\documentclass[12pt,a4paper]{article}
\usepackage{graphicx}
\usepackage{times}
\usepackage{amsmath,amssymb}
\usepackage{natbib}
\usepackage{url}

\title{\bf Radiation-MHD models of elephant trunks and globules in
  H~II regions}
\author{Jonathan Mackey$^{1,2}$\thanks{jmackey@astro.uni-bonn.de} \ and
  Andrew J.\ Lim$^1$\\
\vspace{1cm}\\
\normalsize $^1$ Dublin Institute for Advanced Studies, 31 Fitzwilliam
Place, Dublin 2, Ireland \\ 
\normalsize $^2$ Argelander-Institut f\"ur Astronomie, Auf dem H\"ugel
71, 53121 Bonn, Germany \\
}
\date{\mbox{}}
\begin{document}
\maketitle
\pagestyle{empty}
%
%
\def\bull{\vrule height .9ex width .8ex depth -.1ex}
\makeatletter
\def\ps@plain{\let\@mkboth\gobbletwo
\def\@oddhead{}\def\@oddfoot{\hfil\tiny\bull\quad
``The multi-wavelength view of hot, massive stars''; 39$^{\rm th}$ Li\`ege Int.\ Astroph.\ Coll., 12-16 July 2010 \quad\bull}%
\def\@evenhead{}\let\@evenfoot\@oddfoot}
\makeatother
%
%
\def\beginrefer{\section*{References}%
\begin{quotation}\mbox{}\par}
\def\refer#1\par{{\setlength{\parindent}{-\leftmargin}\indent#1\par}}
\def\endrefer{\end{quotation}}
%
%
{\noindent\small{\bf Abstract:} We study the formation and evolution
  of pillars of dense gas, known as elephant trunks, at the boundaries
  of H~\textsc{II} regions, formed by shadowing of ionising radiation by dense
  clumps.  The effects of magnetic fields on this process are
  investigated using 3D radiation-magnetohydrodynamics simulations.
  For a simulation in which an initially uniform magnetic field of
  strength $\vert \mathbf{B}\vert\simeq50\,\mu$G is oriented perpendicular
  to the radiation propagation direction, the field is swept into
  alignment with the pillar during its dynamical evolution, in
  agreement with observations of the ``Pillars of Creation'' in M16,
  and of some cometary globules.  This effect is significantly enhanced
  when the simulation is re-run with a weaker field of
  $\simeq18\,\mu$G.  A stronger field with $\vert\mathbf{B}\vert\simeq
  160\,\mu$G is sufficient to prevent this evolution completely, also
  significantly affecting the photoionisation process.
  Using a larger simulation domain it is seen that the pillar formation
  models studied in~\citet{MacLim10} ultimately evolve to cometary
  structures in the absence of dense gas further from the star.  }
%
%
\section{Introduction} 
The interstellar magnetic field in the Eagle Nebula (M16) was measured
by~\citet{SugWatTamEA07} using near infrared polarisation observations
of background stars.  They found that the field orientation within the
massive pillars of gas and dust in M16~\citep{HesScoSanEA96}, known as
the ``Pillars of Creation'', is aligned with the long axis of the
pillars and with the UV radiation propagation direction, but
misaligned with the ambient magnetic field (see Fig.~\ref{fig_3}
below).  If the pillars have formed dynamically due to shadowing of
ionising radiation~\citep*[e.g.][]{WilWarWhi01, LimMel03} then it
appears the field has been re-oriented by this process.
\citet{SugWatTamEA07} suggest that this may constrain the ambient
field strength in M16 because ionised gas pressure appears to dominate
the dynamics.  Here we test this suggestion and build on our earlier
non-magnetised results~\citep[][hereafter ML10]{MacLim10} by using 3D
radiation-magnetohydrodynamics (R-MHD) simulations to model the
formation of pillars and globules due to shadowing of ionising
radiation by pre-existing dense clumps.

A uniform grid, 2nd order accurate, finite volume code (see ML10) is
used for the MHD calculations with a Roe-type Riemann solver~\citep{CarGal97}.
The short characteristics ray-tracer is used to track direct
monochromatic ionising radiation and the on-the-spot approximation for
diffuse radiation.  Radiative cooling is calculated explicitly for
recombining hydrogen and using a cooling curve for other elements,
with neutral/molecular gas cooling treated very approximately by
exponential cooling with cooling time $t_c=10\,$kyr (see ML10, cooling
model C2).  This code has been extensively tested\footnote{For more
  information, including test results, see
  \url{http://www.astro.uni-bonn.de/~jmackey/jmac/}. } and
checked for consistency with previous work
(e.g.~\citealt{FalKomJoa98,LimMel03,MelIliAlvEA06,KruStoGar07,HenArtDeCEA09}, ML10
and references therein).

\begin{table}
  \caption{Simulation parameters for model R5, showing grid resolution
    (zones), domain coordinates in parsecs
    $(\mathbf{X}_{\mathrm{min}},\mathbf{X}_{\mathrm{max}})$ relative to the radiation source
    which has monochromatic photon luminosity and photon energy as
    indicated.  Clump 
    positions, peak number density ($n_{\mathrm{H}}$), Gaussian scale radius ($r_0$)
    and total mass ($M$) are
    as indicated; the background magnetic field (in $\mu$G), number
    density ($n_{\mathrm{H}}$) and
    gas pressure ($p_g$) are similarly indicated.\label{tab:R5sim}} 
  \small
\begin{center} 
\begin{tabular}{| l | c  c  c | l  |}
    \hline
    Name & $x$ & $y$ & $z$ & Further information. \\ 
    \hline
    Zones & 384 & 256 & 256 & Uniform Cartesian grid. \\
     $\mathbf{X}_{\mathrm{min}}$ & 1.5 & -1.5 & -1.5 & position in
     parsecs relative to the source. \\
     $\mathbf{X}_{\mathrm{max}}$ & 6.0 &  1.5 &  1.5 & position in
     parsecs relative to the source. \\
     Source &0&0&0& $L_{\gamma}=2\times10^{50}\,\mathrm{s}^{-1}$, $h\nu_0-13.6\,\mathrm{eV}=5.0\,$eV \\
    \hline
     Clump 1 & 2.30 & 0    &   0     & $n_{\mathrm{H}}=10^5\,\mathrm{cm}^{-3}$,
     $r_0=0.09\,$pc, $M=28.4\,\mathrm{M}_{\odot}$ \\
     Clump 2 & 2.75 & 0    &   0.12  & $n_{\mathrm{H}}=10^5\,\mathrm{cm}^{-3}$,
     $r_0=0.09\,$pc, $M=28.4\,\mathrm{M}_{\odot}$ \\
     Clump 3 & 3.20 & 0    &  -0.12  & $n_{\mathrm{H}}=10^5\,\mathrm{cm}^{-3}$,
     $r_0=0.09\,$pc, $M=28.4\,\mathrm{M}_{\odot}$ \\
    \hline
     Background &$\mathbf{B}_0=[0,$ &$0,$ &$53]$ & $n_{\mathrm{H}}=200\,\mathrm{cm}^{-3}$, $p_g=1.38\times10^{-11}
     \,\mathrm{dyne}\,\mathrm{cm}^{-2}$ \\
    \hline
\end{tabular} 
\end{center} 
\end{table} 

\begin{figure}
\centering
\includegraphics[width=1.0\textwidth]{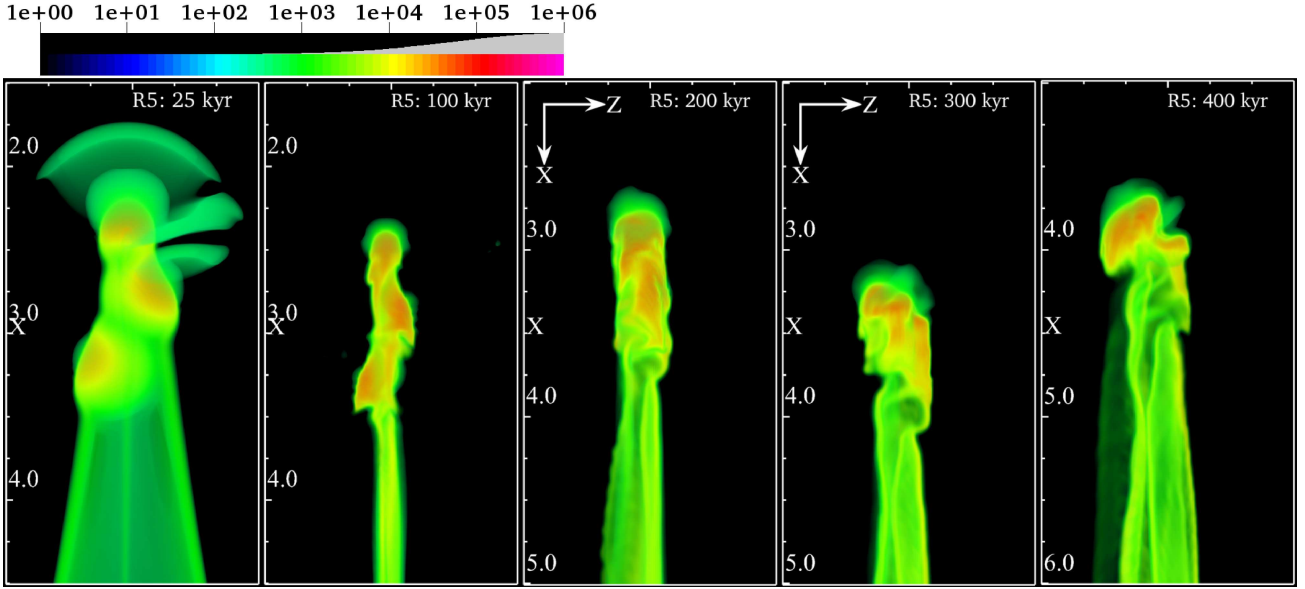}
\caption{Volume-rendered images of gas density in the 3D R-MHD
  simulation R5 described above, shown at times $t= 25,\, 100,\,
  200,\, 300,\, 400\,$kyr.  Tick-marks at $0.25\,$pc intervals show the
  physical scale (numbers refer to the tick-marks immediately below
  them).  The logarithmic number density scale ($\mathrm{cm}^{-3}$) is
  shown above, with the opacity function used for image generation as
  the grey curve (gas with $n_{\mathrm{H}}\lesssim
  10^3\,\mathrm{cm}^{-3}$ is transparent).  Only part of the
  simulation domain is shown in each panel.\label{fig_1}}
\end{figure}

\begin{figure}
  \centering
  \includegraphics[width=0.9\textwidth]{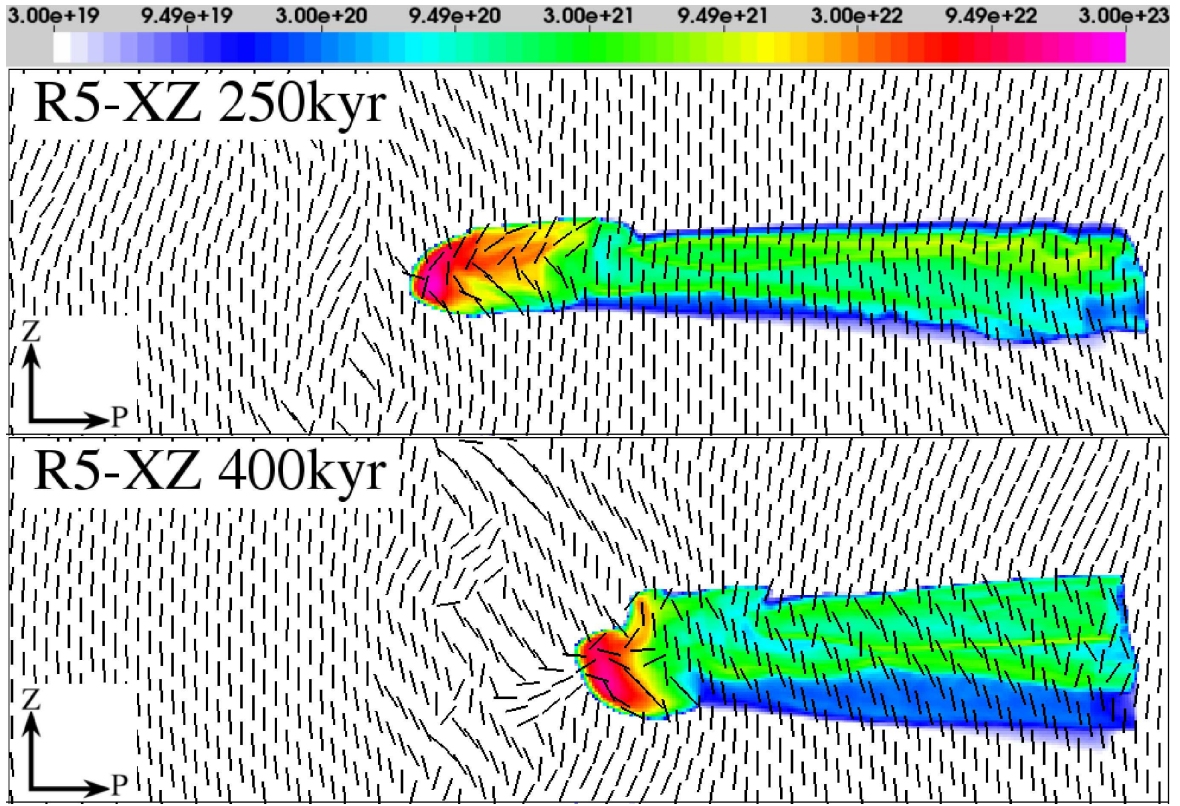}
  \caption{Column density (colour scale in $\mathrm{cm}^{-2}$) and
    projected magnetic field (lines show orientation only) for the R5
    simulation at $250\,$kyr (above) and $400\,$kyr (below).  The
    projection LOS is
    $\hat{n}=\hat{x}\sin(20^{\circ})+\hat{y}\cos(20^{\circ})$ in the
    simulation coordinate system, so the initial magnetic field
    ($\mathbf{B}=53\hat{z}\,\mu$G) is vertical and $\perp$ to the LOS.
    The horizontal axis labelled $\mathbf{P}$ is therefore
    $\hat{P}=\hat{x}\cos(20^{\circ})-\hat{y}\sin(20^{\circ})$ i.e.~a
    $20^{\circ}$ rotation from the $x$-axis about the
    $z$-axis.\label{fig_2}}
\end{figure}

\section{Simulations}
We have performed a number of 3D R-MHD simulations, to
be presented in more detail in~\citet{MacLim11}.  Here we present
results from one of the simulations, denoted R5.  This model consists
of a monochromatic point source of ionising photons placed $2.3\,$pc
from the nearest of three dense gas clumps (Gaussian density profiles)
in a uniform background medium ($n_{\mathrm{H}}=200\,\mathrm{cm}^{-3}$).  Simulation
parameters are described in Table~\ref{tab:R5sim}; the same model was
presented on a smaller simulation domain with zero field in~ML10 but
here a uniform magnetic field of $\mathbf{B}=[0,0,53]\,\mu$G is
imposed in the initial conditions.
 
The evolution of dense gas over $400\,$kyr is shown in
Fig.~\ref{fig_1}, where the initial field is horizontal and in the
plane of the images.  The evolution is broadly similar to that of the R-HD
model in ML10 for the first $\sim250\,$kyr, but there are notable
differences.  Radiation-driven implosion ($t\sim100\,$kyr) is more
pronounced along field lines than across them, leading to a flattened
structure (cf.~\citealt{HenArtDeCEA09}); the narrow tail in the second
panel is actually a sheet seen edge-on and the clumps are also
flattened.  Following this ($200-300\,$kyr) magnetic pressure
generates a stronger re-expansion and hence the pillar/globule is
larger and has a lower density than the R-HD case.  The re-expansion
leads to fragmentation in this model ($t\gtrsim400\,$kyr), just
beginning in the right-most panel.  It is seen that the simulation
ultimately evolves from being pillar-like (dense elongated structure)
to a cometary globule (dense flattened head, low density neutral tail)
in the absence of dense gas further from the star.  We have confirmed
that this is also true when the model is run with zero magnetic field
(i.e.~identical to model 17 in ML10 but with a larger simulation domain).

\section{Discussion}
\begin{figure}
\begin{minipage}{7.5cm}
\centering
\includegraphics[width=7.5cm]{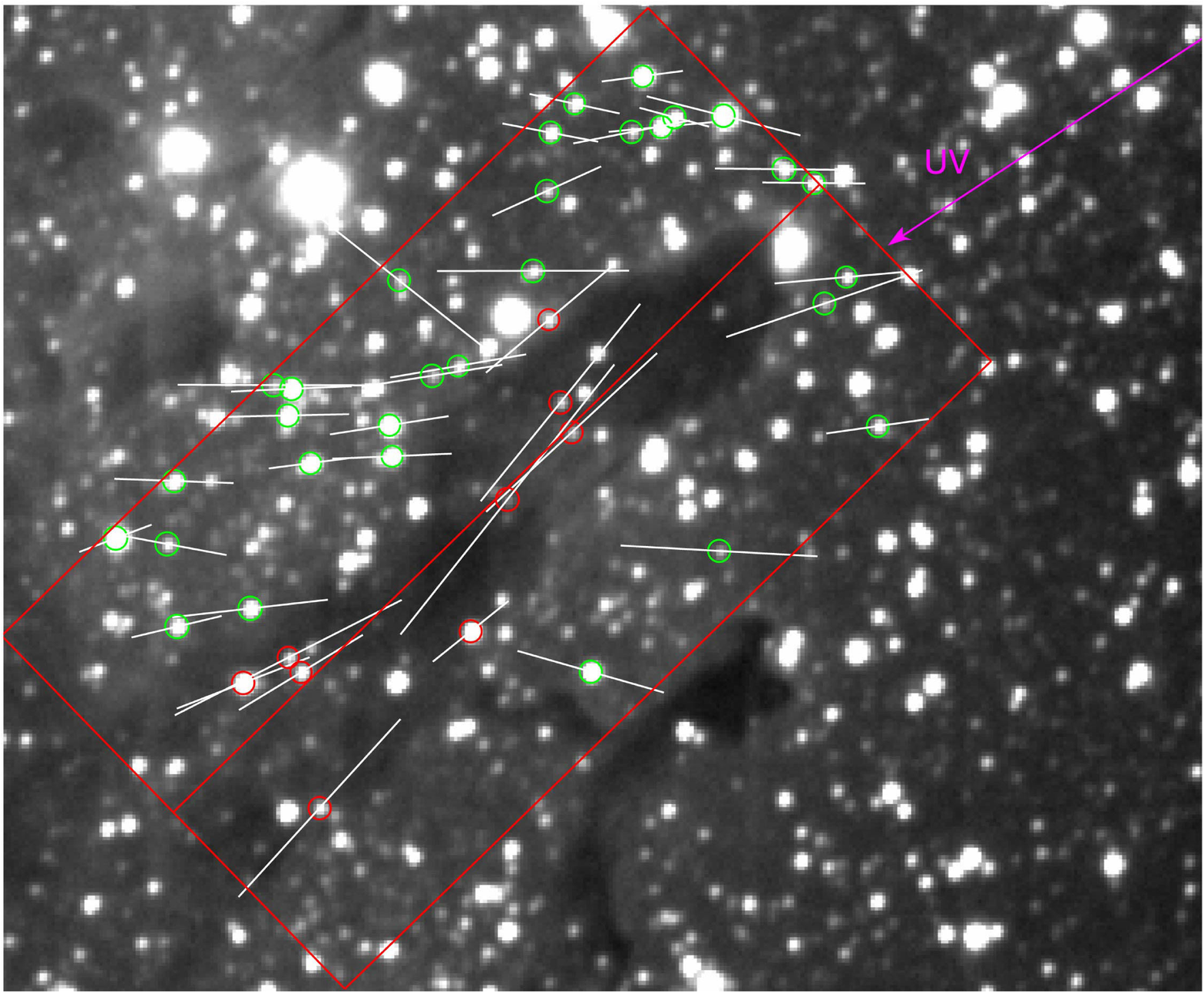}
\caption{Figure 9b from~\citet{SugWatTamEA07} showing near-IR
    absorption polarimetry observations of the magnetic field
    orientation in the central pillar in M16 (\copyright\ the
    Astronomical Society of Japan; used with permission).
    North is to the top, East to the left, and the image width is
    $\simeq2^{\prime}.5$.\label{fig_3}}
\end{minipage}
\hfill
\begin{minipage}{8.5cm}
\centering
\includegraphics[width=8.5cm]{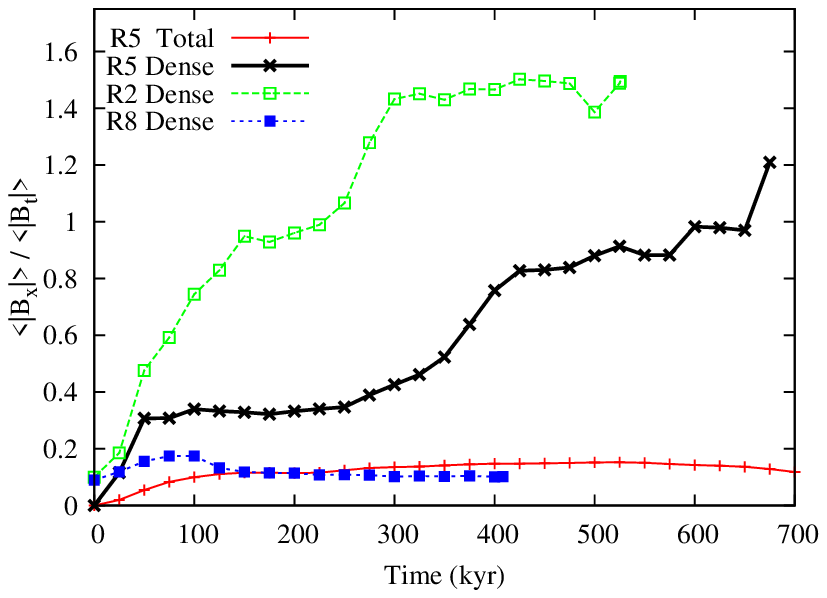}
\caption{Ratio of volume-averaged parallel field $\langle\left\vert
    B_x\right\vert\rangle$ to perpendicular field
  $\langle\sqrt{B_y^2+B_z^2}\rangle$ as a function of time in
  simulations R2, R5, and R8.  Red line: average of all cells (R5);
  Black: average of only dense cells with $n_{\mathrm{H}}>5000\,\mathrm{cm}^{-3}$
  (R5); Green: simulated with $|\mathbf{B}|\ 3\times$ weaker (R2); Blue:
  with $|\mathbf{B}|\ 3\times$ stronger (R8).\label{fig_4}}
\end{minipage}
\end{figure}

The projected magnetic field orientation was calculated by a
density--weighted integration of ``Stokes parameters'' $Q$ and $U$
for the perpendicular magnetic field along the
LOS \citep[cf.][]{ArtHenMelEA10}:
\begin{equation}
  \langle Q \rangle = \int_{z=0}^{\infty}
  \min[n_{\mathrm{H}}(z),n_{\mathrm{max}}]\frac{B_x^2-B_y^2}{\sqrt{B_x^2+B_y^2}}
  dz \,,\quad
  \langle U \rangle = \int_{z=0}^{\infty}
  \min[n_{\mathrm{H}}(z),n_{\mathrm{max}}]\frac{2B_xB_y}{\sqrt{B_x^2+B_y^2}}
  dz \,,
\end{equation}
with $n_{\mathrm{max}}=2.5\times10^4\,\mathrm{cm}^{-3}$.  Here $x-y$
is the image plane and $z$ is distance along the LOS (i.e.~observer
rather than simulation coordinates).  The projected field is then
recovered from $\langle Q \rangle$ and $\langle U \rangle$ using
trigonometric relations.
Formally this has
units of $\mu\mathrm{G}\,\mathrm{cm}^{-2}$, but the normalisation is
irrelevant for the orientation.  The projected field orientation is
over-plotted on column density maps in Fig.~\ref{fig_2}, projected
such that the initial field is fully in the image plane and vertical.
At $t=250\,$kyr the clumps have undergone significant dynamical
evolution and the pillar-like structure is slowly being flattened due
to the rocket effect.  The magnetic field, while relatively unaffected
in ionised gas and the low density tail region, has clearly been
dragged into alignment with the pillar in dense gas due to the
pillar's acceleration away from the star.  At $400\,$kyr only a
flattened cometary globule remains, but the field is even more
strongly aligned with the acceleration direction.  This clearly shows
that an initially perpendicular field is swept into alignment with the
pillar during its dynamical evolution, and that this alignment remains
once the structure evolves to a cometary morphology.  This is in
agreement with the observations of the pillars in M16 (reproduced here
in Fig.~\ref{fig_3},~\citealt{SugWatTamEA07}) and some cometary
globules~\citep[e.g.][]{BhaMahMan04} which also have the field aligned
with the cometary tail.

The behaviour of the same simulation with a magnetic field $3\times$
weaker (R2) and $3\times$ stronger (R8) is rather different
however~\citep[to be discussed in more detail in][]{MacLim11}.  For R2
with $B\simeq18\,\mu$G the field is more easily deformed by
photoionisation--induced gas motions and the alignment is even
stronger, whereas for R8 with $B\simeq160\,\mu$G the field is
dynamically dominant and barely changes from its initial perpendicular
state.  This is shown in the right panel of Fig.~\ref{fig_3} where the
ratio of the volume averaged field parallel and perpendicular to the
radiation propagation direction ($\hat{x}$) are plotted as a function
of time for R2, R5, and R8.  The evolution of dense gas with
$n_{\mathrm{H}}>5000\,\mathrm{cm}^{-3}$ is shown for all three models,
and for the full simulation volume only for R5 (R2 and R8 are
very similar).  It is seen that while the ratio does not evolve
significantly when averaged over the full simulation volume, in dense
gas the situation is very different.  For R5 the parallel field
increases rapidly during the implosion phase and there is a subsequent
slower increase during the acceleration phase to roughly equal field
strength in both components.  This trend is stronger for the weak
field model R2, whereas for the strong field model R8 the field
orientation remains almost constant throughout the simulation.  This
difference between the weaker and stronger field simulations 
is larger than was
found by~\citet{HenArtDeCEA09} and deserves further investigation.

\section{Conclusions}
The results presented here show that both radiation-driven implosion
and acceleration of clumps by the rocket effect tend to align the
magnetic field with the radiation propagation direction in dense
neutral gas.  As was suggested by~\citet{SugWatTamEA07}, the
effectiveness of this alignment is dependent on the initial field
strength.  For simulations which have similar gas densities and
pressures to conditions in M16, we have shown that a magnetic field of
strength $\sim160\,\mu$G is sufficient to prevent any significant
field reorientation.  These results suggest that an ambient field of
$\vert\mathbf{B}\vert<160\,\mu$G (and more likely
$\vert\mathbf{B}\vert\sim50\,\mu$G) is required to explain the
observed field configuration in the M16 pillars if the pillars formed
via the mechanism we are modelling.  Detailed R-HD models have
recently been performed with dynamic initial
conditions~\citep{GriBurNaaEA10} rather than the initially static
models of ML10, allowing a potentially more realistic comparison with
observations.  Addition of magnetic fields to their simulations would
be very useful to assess how significantly a non-uniform initial
magnetic field will impact on the results presented here.

%
%
\section*{Acknowledgements}
JM acknowledges funding for this work from the Irish Research Council
for Science, Engineering and Technology; from the Dublin Institute for
Advanced Studies; and from Science Foundation Ireland.  AJL's work has
been funded by a Schr\"{o}dinger Fellowship from the Dublin Institute
for Advanced Studies.  The authors wish to acknowledge the SFI/HEA
Irish Centre for High-End Computing (ICHEC) for the provision of
computational facilities and support.  Fig.~9b
from~\citet{SugWatTamEA07} \copyright\ The Astronomical Society of
Japan; reproduced with permission.
%
%

{\footnotesize
\bibliography{refs}
}
\end{document}